\documentstyle[psfig,conf_iap,]{article}
\begin{document}
\heading{%
%Begin Heading
%
Chemical Abundances in High-redshift Neutral Clouds 
 
%
%End Heading
} 
\par\medskip\noindent
\author{%
%Begin Author names
Paolo Molaro$^{1}$
%End Author names
}
\address{%
%First address
Osservatorio Astronomico di Trieste,
Via G.B. Tiepolo 11,
I-34100 Trieste, Italy.
}

\begin{abstract}
Neutral hydrogen clouds with high column density 
detected  towards
distant quasars are unique
probes of elemental nucleosynthesis and chemical evolution in the  
low  metallicity regime.
They   provide measurements for several elements 
at very early times 
which are unfeasible in other 
astrophysical environments. Comparison between refractory and non-refractory elements 
provides evidence for  the presence of dust, and the recently measured Ar probes 
photoionization.
A prominent  characteristic 
is the dominance of a solar abundance pattern, which is 
 somewhat unexpected at low  metallicities. It is argued that this property and 
Nitrogen observations can be used to 
constrain the age  of the  Damped Ly$\alpha$ systems and the 
epoch of star formation.

\end{abstract}
\section{Introduction}
Quasar absorption systems characterized by
 a large hydrogen column density, $\log N(HI) \ge$  20.2 cm$^{-2}$
  (hereinafter DLAs), are  
 observed at all redshifts 
up to z $\approx$ 4.5 (Dessaugues-Zavadsky et al 2001) 
although there is some evidence
for a  lack of 
large HI column density
systems for z$\ge$ 4 (Peroux et al 2001). DLAs are the largest reservoir
 of neutral hydrogen and there is little doubt that they 
  are the
 progenitors of the present day galaxies.
However, the main galactic  population  which originates   the DLAs
 has still to be clearly identified.
  At low redshift deep imaging has revealed  a heterogeneous
population of   dwarfs, low surface brightness, irregulars and also spirals
galaxies (Le Brun et al 1997, Rao \& Turnshek 2000).  
%At low redshift IZw18 or the Milky Way itself 
%will be classified as damped Ly$\alpha$ systems according to the value of 
%their neutral hydrogen column densities. 

Some sort of  chemical information 
is available for  almost half of the $\approx$  180 DLAs  presently 
known. Frequently 
 measured 
ions  are   
  SiII, FeII, CrII,  ZnII, 
 NiII, AlII,  MnII, TiII, while there is relatively less   information 
  for elements
 such as SII, NI  ArI, PII,   
 and OI whose useful 
 transitions are mixed and often blended  by  the intervening Lyman forest.
 Recently,  the D/H ratio has also been successfully measured 
 in a DLA (Levshakov et al 2002).
 Abundances   in the DLAs can be rather  accurate with 
errors  of the order of  25\% - 10\%,   or even better 
in case
of optically thin lines and  simple velocity structure.
There are no other techniques applied to  the high redshift universe
where chemical abundances can be obtained with comparable  precision.
However,  different elements show 
 different sensitivities to the radiation field, 
or  affinity 
with dust, which    somewhat confuses the  nucleosynthetic pattern and 
complicates
the interpretation of the observed abundances.

\section{Dust, Argon and ionization}

 The presence of dust in the clouds
 affects   the abundance of those  elements which are locked up 
 into  dust grains
    and are therefore depleted 
 from the diffuse phase. Several indications 
 suggest  the presence of dust in  DLAs. Pettini et al (1997) 
 pointed out that Fe and Cr are generally deficient in comparison  to
 the  undepleted  Zn, as  is found in the dusty Galactic disk and
   at variance with what is observed
 in the metal poor stars of the Galactic halo  where 
 Zn  tracks  Fe  closely. 
 We note here that  in  DLAs iron  is always found slightly 
 more deficient  than Cr,
 which may indicate the presence    of a differential
 depletion of the two elements as it also found 
 in warm interstellar clouds of our own Galaxy (Savage and Sembach 1996).
Hou et al  (2001) found that 
 the abundances of refractory 
 elements  anti-correlate  
  with the elemental column density, which mimics  the
 well  known pattern  between these  
  elements and  the hydrogen column density observed in the Milky Way.
 %(Wakker and Mathis 2000). 
 New evidence for the presence of dust
 is provided by the few systems 
 where molecular hydrogen has been detected.
 Levshakov et al (2002) found that the fraction of  
   molecular hydrogen increases in the DLAs showing more 
  dust-depletion inferred from the [Fe/Zn] ratios,
   as  expected
  since   dust is needed in 
  the process of H$_2$ 
 formation. 
 
 The large hydrogen column densities 
shield efficiently the clouds from the surrounding IGM 
radiation field (Viegas 1995). 
Thus,   
 DLAs are  essentially HI regions 
with  the lower ionization elements   as the 
dominant state. High ionization species such as 
CIV and SiIV  are often 
 observed but show  different velocity structure and 
  originate quite likely in different regions.  
Recently it has been pointed out that AlIII and AlII 
have  similar profiles  
suggesting the presence of HII interfaces.
% (Prochaska \& Wolfe 2000, Howk \& Semback 2000).  
However, Vladilo et al (2001)  showed  that even when 
such interfaces are accounted for,  the  corrections remain  
negligible for all elements with the possible exception of 
but possibly Al.
A radically different model where the  
 DLAs are an ensemble of neutral metal-free gas and
 ionized gas containing metals 
 has been proposed by 
 Izotov et al (2001).  
  These  models foresee significant ionization and 
can be tested with  
  Argon, which has been recently measured in    
  a couple of DLAs (Molaro et al 2001; Levshakov et al 2002).
 Argon has a photoionization cross-section which is
$\ge$ 10 times higher than of HI and therefore 
is very sensitive to ionization effects 
(Sofia and Jenkins 1998). 
Space observations 
showed that in the LISM the (Ar/H) gas-phase is below solar 
by $-$0.3 to $-$0.6 dex varying with the line of sight.
Argon is not expected to be depleted in the interstellar medium and the
observed deficiency  has been interpreted by Jenkins and Sofia 
as due to partial 
ionization by UV stellar radiation.
The fact that the Ar  abundances  in the two DLAs  where Ar has been measured
are almost  in line 
  with   the  
   abundances of O, S and  Si,  requires  that ionization effects 
  are relatively unimportant. In fact, by applying 
the Izotov et al correction to Ar would result in an Ar  
    
  Ar overabundances  by one dex or more, which is quite unlikely.
  At the same time these measures imply that in this DLA 
   no intense star formation 
 is  taking place along the line of sight. 
 New measures of Ar in other DLAs 
  would be of great importance for probing general ionization effects 
  either due to the intergalactic radiation
   field or to star forming regions inside the clouds themselves.

\section{The chemical pattern and the (S/O), (Ar/O) ratios}

Once the effects of ionization and dust are considered  it 
is possible  to address the isue of the presence of
 a specific  chemical pattern  
for  the DLAs. The ratio between $\alpha$ 
elements and the iron-peak elements, which is related to the relative
 contribution
of Type II and Type I supernovae, 
is a key piece of information for  tracing back the 
chemical history.  
The early works focussed on  the fact that  the  [Si/Fe] ratio
is  $\approx$  0.4 on average, as   is  
 observed in the
Galactic halo stars, and  this  was  readily 
interpreted  as the signature of  TypeII SN. 
\begin{figure}
\centerline{\vbox{
\psfig{figure=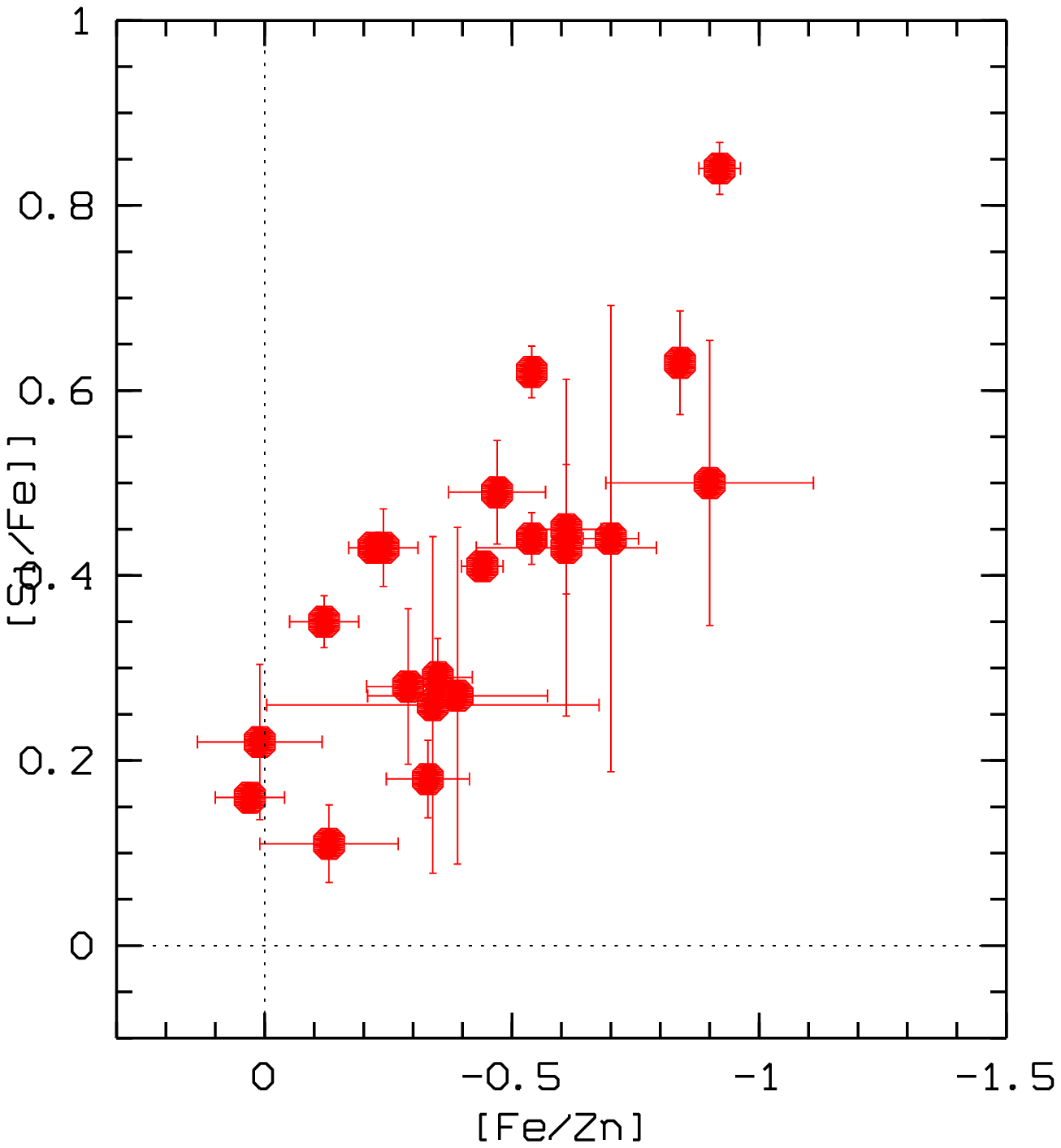,height=6.cm}
}}
\caption[]{ [Si/Fe] versus [Fe/Zn] }
\end{figure}
However, a  similar ratio  is observed also in the warm phase gas
of the LISM, 
and  the former   interpretation is valid only 
in  complete absence of dust, which we regard as  unlikely.
Fig 1  shows      the [Si/Fe] versus   the 
[Fe/Zn]  ratios. The strong correlation between the two quantities
 is very  suggestive of the presence of differential  depletion 
of Fe relatively to both Si and  Zn.   The intercept 
for [Fe/Zn]=0, gives [Si/Fe] $\approx$ 0.1, providing evidence for only 
a mild enhancement of the [Si/Fe] ratios in the less dusty clouds.
It is also rather interesting to note  that 
both  the behaviour and the range of the [Si/Fe] and [Fe/Zn] ratios 
in the DLAs  are similar to those  observed  in  
 isolated Galactic 
clouds 
presented  by  Lew Hobbs in  this conference.

Different    approaches  have been followed to resolve 
 the degeneracy   in the interpretation
 of the [Si/Fe] ratio (Molaro 2002).
 Some  authors have tried to correct for dust (Vladilo 1998), 
 or considered only systems
 showing little evidence for dust (Pettini et al 1999) 
 or considered only non-refractory elements such as the [S/Zn] (Centurion et al 2000).
    The picture  that is emerging  favours a variety of chemical patterns with 
 both halo-like and solar chemical patterns
  present
 in  DLAs. The latter, which are somewhat unexpected on theoretical grounds
 are in fact those  
   more frequently found.
   Solar ratio at low metallicity requires that the
 yields of SNe Ia  have already become effective in determining the global 
 metallicity of the system. This can be achieved either with a slow SFR or from a burst
 of SF followed by a long quiescent period in which  SNe Ia  evolve
   (Matteucci et al 1997). If the relatively modest  $\alpha$ enhancement 
   is due to the action of 
 SNe Ia, there are at least two important implications.
 The first is that  SNe Ia are not suppressed at low metallicity
 as argued by Nomoto 
 and collaborators. The second is that 
the epoch of star formation  in the systems has to be 
anticipated to cope at least with  SNe Ia lifetimes and subsequent 
elemental mixing.
   
 %However,
 %in an instantaneous   burst of star formation, the timescales for  type Ia SN 
%ay be considerably short as $\approx$ 40-50 Myr (Matteucci \& Recchi 2001)
 
\begin{figure}
\centerline{\vbox{\psfig{figure=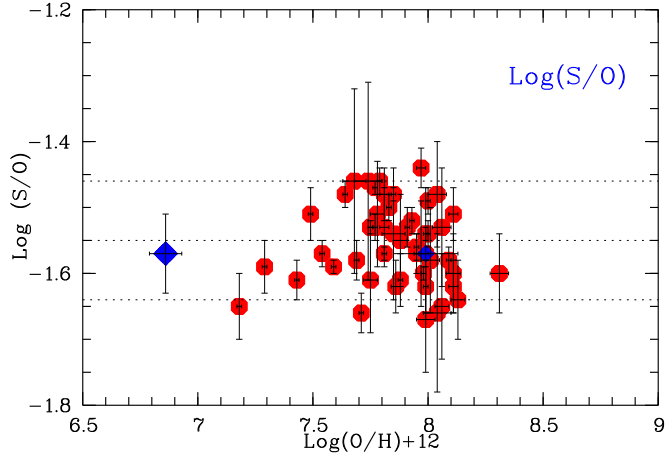,height=6.cm,angle=-90}
}}
\caption[]{The $\log$(S/O) in the DLAs (diamonds).
 The small exagons are  the BCGs
from Izotov and Thuan (1999), with the mean dispersion shown by the dotted line}
\end{figure}

Henry and Worthey (1999) pointed out that the observed ratios 
of (S/O)  and of (Ar/O) are  found to be  independent from metallicity
in a large variety of objects including the Milky-Way,  HII spirals
and  Blue Compact Galaxies (BCG).
New UVES-VLT observations provided for the first time 
information for these three elements so that these  abundances
can be discussed in the same context.
The $\log (S/O)$  ratios in the DLAs, shown together 
with those of the BCGs in Fig. 2,  taken from 
 Izotov \& Thuan 
(1999) are  indistinguishable from the BCGs, and the  same occurs 
for the $\log (Ar/O) $.
 %The average value of the 
%BCGs  is $\log$(Ar/O)= --2.25 $\pm$0.09, 
%and $\log$(S/O)= --1.55 $\pm$0.09, which
%is also close to   theoretical model predictions 
%( $\log$(Ar/O)= $\approx$ --2.5 $\log$(S/O) 
%= $\approx$ --1.7 
%(Woosley \& Weaver 95, Nomoto et al '97, 
%Samland et al '98).
The universal constancy  of the $\log (S/O)$ and $\log (Ar/O)$ ratios 
found in different galaxies is 
extented now also to the   DLAs, at least in the few cases
 where such measures 
have become available,
and probed  to 
an even  lower value of $\log (O/H) + 12$. 
This is particularly 
significant  considering that DLAs  and BCGs in 
the local universe
may have formed at totally different epochs.
 Since  the O yields are more 
sensitive to  progenitor mass than S and Ar, the constancy 
in the ratios between these $\alpha$ elements has been interpreted by
Henry \& Worthey (1999) as  evidence for a  universal 
IMF in different galaxies, and this argument can be now developed 
to include also the   DLAs.

\section{Special elements:  Phosphorus \&  Nitrogen}

Phosphorus is a recent entry in the set of elements measured in the DLAs
with  3 measures reported 
in the systems  towards QSO 0000$-$2621 (Molaro et al 2001)
 GB1759+7539 (Outram et al 1999) and QSO 0347$-$3819 (Levshakov et al 2002).
 Phosphorus   cannot be measured in halo stars
and   DLAs offer a unique site  where it
 can be measured at  metallicities significantly lower than  solar.
 The  $^{31}$P is an  odd-element which is 
 likely produced by $n$-capture in C and Ne shell burning.
In the dust free DLA  towards QSO 0000$-$2621 phosphorus  is found 
slightly below iron,  
$[P/Fe] = -$0.2, 
which requires a sort of mildly   metallicity dependent yields to arrive 
at solar values later on.
The relative ratios of P with the closeby nuclei S and Si,  
$[P/Si,S] = -$0.3,   show evidence for a mild odd-even effect, but it is somewhat 
lower than what is predicted by the theoretical models of Limongi et al (2000).
P is not depleted in the LISM and, once the nucleosynthetic properties 
of this element are better defined with 
additional observations,
it
 may  become a useful proxy for Zn and 
 a potential tracer of the metallicities 
 for very high redshift damped systems where Zn is difficult to detect.

\begin{figure}
\centerline{\vbox{\psfig{figure=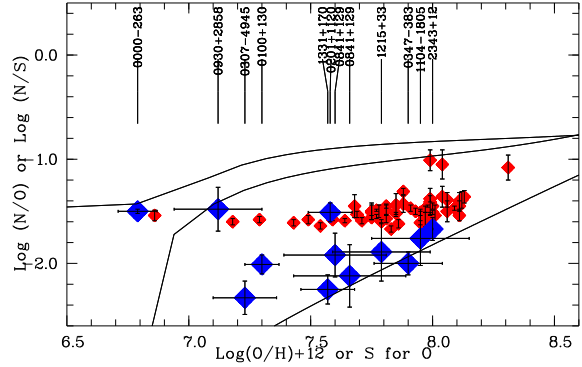,height=7.cm,angle=-90}
}}
\caption[]{Curves are
for secondary production only (bottom),
primary production in 1$<$ M$_{\odot}$$<$8, and primary for  M$_{\odot}$$>$8 
(top).
Data points are from 
Centurion et al (1998 and unpublished), Ellison et al (2001),   
Dessauges-Zavadsky et al (2001). The small diamonds are the  
 BCGs  from Izotov \& Thuan (1999)}
%and Kobulnicky and Skillman (1996)}
\end{figure}
 
The available Nitrogen measures in the DLAs are 
shown in Fig. 3 where the $\log (N/O)$ versus $\log O + 12$ are plotted.
The N observations are taken from different 
 sources as described in the  caption. O is directly 
  measured in only 3 DLAs 
towards QSO 0000$-$2621, BR J0307$-$4945 and  QSO 0347$-$3819. So that for the 
remaining ones we use 
 S  as a proxy for O.  
 Nitrogen has a rather complex nucleosynthesis.
Its production occurs mainly in intermediate mass stars with 
4$-$8 M$_{\odot}$, which undergo HB burning 
and expel large amounts of primary N at low [Fe/H] and 
secondary-tertiary N at higher metallicities.  
A primary production at low
 metallicities by massive stars with M$>$ 8 M$_{\odot}$  has also been 
 suggested. The theoretical curves corresponding to
 these  different scenarios are shown in the figure, together with 
the  measurements in the Blue Compact Galaxies by  Izotov and Thuan (1999).
 As it is possible to see from the figure
 the  DLAs data points  show values 
 close to the  theoretical curve for  secondary production
and other values   in correspondence of the theoretical curve 
for   primary behaviour.  In particular, 
 in some DLAs 
 $\log (N/O)$  is  much   lower  than any other measurement in the BCGs.
Izotov et al argued that  ionization effects may lower  the  NI/SII ratios,   
and therefore  the $N/O$, if  S is 
 taken as a proxy for O. However, the new direct oxygen 
 measurements show that this is not
 the case since the argument  does not apply to   OI. 
The low  $\log (N/O)$=$-$2.4 in BR J0307$-$4945
  obtained from a  direct   measure of O 
shows that   low values of $\log (N/O)$
are real and provides some evidence against the possibility
of a primary origin of N in
 massive stars.

The observed scatter 
 can be interpreted as due to  the  delayed release  of N  relatively to O.
 For instance 
only 6 Myrs are required for a   
25 M$_{\odot}$ star to release O, while    N requires longer 
timescales, i.e.   about  250 Myrs for 
a   4 M$_{\odot}$ star.
Within this  model   the $\log (N/O)$
measurements offer the  way to
 constraint the    ages of the clouds. The  low-$\log (N/O)$ cases 
 would correspond to   young objects,
$\le$ 250 Myrs, while 
high $N/O$   to  older systems. 
In  support of this interpretation there is  also  a hint of 
correlation between  the $\log (N/O)$ and the $\alpha$-element enhancement.
DLAs with  low $\alpha$-enhancement such as QSO 0000$-$2621 or QSO 0930+2858 show a tendency
for  high $N/O$ abundances, while    low $N/O$ values are observed 
in   $\alpha$-element enriched DLAs such as the one  towards 
QSO 0347$-$3819. The former two DLAs are at z$_{abs}$$\approx$ 3.4  
and  require  the epoch of star formation 
at  earlier times to  produce both considerable amount of N and to 
make the TypeIa yields effective in reducing the 
$\alpha$ over iron-peak ratios. Assuming an indicative  
time interval of 500 Myrs the star formation is placed at z $\ge$ 5,
for a $\Lambda$CDM cosmology with $\Omega_{\lambda}$=0.7, $\Omega_{m}$ =0.3,
$\Omega_{b}$=0.05 and H$_{0}$= 70 km s$^{-1}$ Mpc$^{-1}$.
Thus it may well be possible that at high redshift some  neutral clouds 
are  rather passive systems chemically enriched by an earlier
stellar population. This would also explain the {\it normal} Ar abundances,
the absence of detectable Ly$\alpha$
flux  and the apparent lack of  chemical evolution of DLAs with redshift.

%   Q1331+1700 and PHL 957 show low N/O AND low $\alpha$.

\acknowledgements{I warmly  acknowledge  Miriam Centurion, Piercarlo Bonifacio,
Sergio Levshakov, Giovanni Vladilo, Sandro D'Odorico 
and Miroslava Dessauges-Zavadsky  for allowing me to present 
 common work and Sara Ellison for useful comments on an early version of the paper.}

\vfill
\end{document}